\documentclass[reprint, amsmath,amssymb,nofootinbib, aps,superscriptaddress,longbibliography]{revtex4-1}
\usepackage[utf8]{inputenc}
\usepackage[english]{babel}
\usepackage[T1]{fontenc}
\usepackage{amsmath}
\usepackage{amssymb}
\usepackage{graphicx}
\usepackage[normalem]{ulem}
\usepackage{float}
\usepackage{comment}
\usepackage{xcolor}
\usepackage{tikz}
\usepackage{lipsum}
\usepackage{hyperref}
\usepackage{cancel}
\usepackage{graphics}

\usepackage{etoolbox}
\usepackage{adjustbox}
\newcommand{\figref}[1]{Fig.~\ref{fig:#1}}

\newcommand{\figrefbegin}[1]{Figure~\ref{fig:#1}}
\newcommand{\tableref}[1]{Table~\ref{tab:#1}}
\newcommand{\secref}[1]{Sec.~\ref{sec:#1}}
\newcommand{\appref}[1]{appendix~\ref{app:#1}}
\renewcommand{\eqref}[1]{Eq.~(\ref{eq:#1})}
\addto\captionsenglish{}

\newcommand{\eqsref}[2]{Eqs.~(\ref{eq:#1},\ref{eq:#2})}

\renewcommand{\vec}[1]{\mathbf{#1}}

\newcommand{\mat}[1]{\mathbb{#1}}

\newcommand{\BRA}[1]{\left<{#1}\right.\!|}
\newcommand{\KET}[1]{|\!\left.{#1}\right>}

\newcommand{\Scale}[2][4]{\scalebox{#1}{$#2$}}

\usepackage{tikz,lipsum,lmodern}
\usepackage[most]{tcolorbox}
\usepackage{etoolbox}
\AtBeginEnvironment{tcolorbox}{\small}

\definecolor{my-blue}{cmyk}{0.80, 0.13, 0.14, 0.04, 1.00}
\definecolor{mypurple}{rgb}{0.59, 0.44, 0.84}

\begin{document}

\title{Inertial geometric quantum logic gates}

\author{D.Turyansky}
\affiliation{Department of Applied Physics, The Hebrew University of Jerusalem, Jerusalem 9190401, Israel}
\author{O. Ovdat}
\affiliation{Department of Chemical Physics, Weizmann Institute of Science, Rehovot 76100, Israel}
\author{R. Dann}
\affiliation{Max-Planck-Institut f{\"u}r Quantenoptik, Hans-Kopfermann-Straße 1, D-85748 Garching, Germany}
\affiliation{The Institute of Chemistry, The Hebrew University of Jerusalem, Jerusalem 9190401, Israel}
\author{Z. Aqua}
\affiliation{Department of Chemical Physics, Weizmann Institute of Science, Rehovot 76100, Israel}
\author{R. Kosloff}
\affiliation{The Institute of Chemistry, The Hebrew University of Jerusalem, Jerusalem 9190401, Israel}
\author{B. Dayan}
\affiliation{Department of Chemical Physics, Weizmann Institute of Science, Rehovot 76100, Israel}
\author{A.Pick}
\affiliation{Department of Applied Physics, The Hebrew University of Jerusalem, Jerusalem 9190401, Israel}
\email{adi.pick@mail.huji.il}

\begin{abstract}
 We present rapid and robust protocols for STIRAP and quantum logic gates.  Our gates are based on geometric phases acquired by instantaneous eigenstates of a \emph{slowly accelerating} ``inertial'' Hamiltonian. To begin, we establish the criteria for inertial evolution and subsequently engineer pulse shapes that fulfill these conditions. These tailored pulses are then used to optimize geometric logic gates.  We analyze a realization of our protocols with $^{87}$Rb atoms, resulting in gate fidelity that approaches the current state-of-the-art, with marked improvements in robustness.
\end{abstract}

\maketitle
\section{Introduction}

High-fidelity quantum logic gates play a critical role in advancing large-scale atom-based quantum computing systems~\cite{ladd2010quantum,saffman2016quantum,henriet2020quantum,morgado2021quantum}. Presently, the record fidelity stands at 99.99$\%$ for single-qubit gates~\cite{sheng2018high,levine2022dispersive}, 99.5$\%$ for two-qubit gates using hyperfine qubits~\cite{levine2019parallel,evered2023high}, and 99$\%$ for qubits combining hyperfine and Rydberg levels~\cite{madjarov2020high}.  Advancing these capabilities remains a primary objective in quantum computation research.  Gate fidelity is limited by both fundamental and technical noise sources~\cite{saffman2010quantum}. Accordingly, protocols are optimized to reduce errors by compromising the different noise sources, e.g., minimizing the population in radiative states, reducing the input power,  and the gate time.   Recent work on Rydberg-based gates suggests that robust pulses,
 designed to withstand Doppler and intensity noise, could surpass existing state-of-the-art gate fidelities. Such pulse optimization is anticipated to yield substantial enhancements in error-correcting codes~\cite{wu2022erasure,jandura2023optimizing}. These developments motivated our search for general principles that produce rapid and robust gate protocols. 

Our approach is a modification of traditional adiabatic following of eigenstates, where a slowly varying system adheres to an instantaneous eigenstate during its evolution~\cite{born1928beweis}. While adiabatic protocols exhibit robustness against noise, their limitation lies in speed. 
Here, we use  \emph{inertial protocols}~\cite{dann2021inertial} -- where the system adheres to eigenstates of an inertial Hamiltonian (defined below) --  which   
match the robustness of adiabatic protocols but offer faster execution. We present  analytic inertial pulse shapes for stimulated rapid adiabatic passage~\cite{vitanov2017stimulated} (STIRAP) and geometric  logic gates~\cite{unanyan1999laser,duan2001geometric,moller2007geometric,toyoda2013realization} (\figref{Fig3}).
Then, we improve their performance with quantum optimal control (QOCT). 
We construct an optimization functional that includes 
penalty terms for deviations from the adiabatic and inertial conditions [\eqsref{inertial-param}{adiabatic-param}]. Finally, we use Krotov's algorithm to carry out the optimization.  
The performance of our protocols is analyzed in \figref{Fig2}, demonstrating gate fidelity close to the state-of-the-art with improved robustness. 
Inertial protocols could potentially improve additional adiabatic protocols, including quantum search ~\cite{daems2008analog},  teleportation~\cite{bacon2009adiabatic},  quantum annealing~\cite{sels2017minimizing}, and additional entangling gate protocols~\cite{saffman2020symmetric,li2022single}.

The paper is structured as follows. We begin by surveying the literature on adiabatic protocols and quantum optimal control (\secref{Background}) and continue by introducing the inertial theorem (\secref{Inertial-thm}). Then, we apply the theorem to find advantageous pulse shapes for STIRAP (\secref{inertial-STIRAP}). To quantify the advantage of our approach, we benchmark several  STIRAP implementations in terms of their fidelity, robustness, and pulse area (\secref{benchmark}). In \secref{QOCT-sec}, we outline our QOCT algorithm, which includes inertial constraints. \secref{Geometric-gates} presents an application of our STIRAP pulses for geometric quantum logic gates. \secref{Noise} and \secref{realization} present an analysis of the dominant noise channels and a proposal for experimental realization with $^{87}$Rd atoms. We conclude with a discussion of the advantages and shortcomings of the present approach (\secref{Conclusion}).

\section{Background}\label{sec:Background}

Many efforts have focused on improving adiabatic protocols, aiming to enhance fidelity within a specified pulse area while preserving their robustness~\cite{vitanov2017stimulated}. Below, we survey some of these significant findings. A straightforward strategy involves seeking pulses that minimize diabatic errors, arising from population leakage from the instantaneous eigenstate.  Diabatic errors can be expressed in terms of complex poles of the energy gap~\cite{davis1976nonadiabatic}.  It follows that protocols that minimize diabatic errors involve parallel transport of instantaneous eigenvalues~\cite{guerin2002optimization}.   Another effective strategy involves fulfilling the adiabatic condition at each moment during the propagation by appropriately adjusting the instantaneous rate of change of control fields in proportion to the instantaneous energy gap~\cite{malinovsky1997simple,roland2002quantum}.   These methods have found success in applications including adiabatic state transfer~\cite{lacour2008optimized,dridi2009ultrafast,vasilev2009optimum}.

Other methods extend beyond merely tracking the instantaneous eigenstates adiabatically. For example, some protocols rely on tracking ``superdiabatic basis states,'' which are successive truncations of the series solution of the  Schrödinger equation in powers of the adiabatic parameter~\cite{lim1991superadiabatic}. Following superdiabatic states is favorable in computation since these states adhere more closely to the physical solution than adiabatic eigenstates.    Another approach for improving adiabatic protocols consists of introducing    ``counter-diabatic'' control fields that cancel diabatic errors~\cite{berry2009transitionless}. Both superdiabatic and counterdiabatic protocols are part of a larger class of methods called shortcut-to-adiabaticity (STA) methods~\cite{guery2019shortcuts,huang2017fast}. In theory, STA  can achieve perfect fidelity in a short time. However, practical bounds on power and speed limit the fidelity in practice. Counter-diabatic driving was successfully applied to STIRAP in~\cite{du2016experimental}. Combining counter-diabatic driving with superdiabatic basis states can produce even faster robust protocols~\cite{benseny2021adiabatic,baksic2016speeding}. Our approach is a part of   STA methods, which involve tracing eigenstates of modified Hamiltonians. We mention an early related proposal for following eigenstates of non-trivial invariant operators~\cite{chen2012engineering}, which is similar in spirit to invariants that arise in the inertial frame.

Adiabatic protocols can be improved with   QOCT~\cite{werschnik2007quantum,glaser2015training,koch2022quantum}.  First, adiabatic protocols can be used as an initial guess for the optimization, which is aimed at increasing gate or state-transfer fidelity~\cite{wang1996optimal,saffman2020symmetric,fu2022high}. To find pulses that not only maximize the fidelity but are also robust, one can optimize an ensemble of protocols that samples realistic fluctuations~\cite{goerz2014robustness}. In Ref.~\cite{jandura2023optimizing}, analytic considerations were used to find robust optimized pulses. In~\cite{mortensen2018fast}, QOCT was used to search for solutions that  closest to the STA  within a set of implementable pulses. 
Alternatively,  the optimization functional can be modified to include constraints for adiabatic following. For example, one can add a constraint to maximize projection onto dark  states~\cite{brif2014exploring}. A related idea is to minimize projection onto radiative  states~\cite{palao2008protecting}. Another approach is to satisfy the adiabatic condition locally~\cite{malinovsky1997simple, sola1999optimal}. Yet another approach applies learning algorithms~\cite{yang2020optimizing}. Specifically,  we mention proposals that are closely related to our work, which use  QOCT to improve STIRAP~\cite{mortensen2018fast,wang1996optimal,malinovsky1997simple,sola1999optimal},  adiabatic Rydberg-based gates~\cite{goerz2014robustness,saffman2020symmetric}, and   non-adiabatic  Rydberg gates~\cite{jandura2023optimizing,fu2022high,zheng2023thermal}.
We developed a QOCT approach that finds protocols that adhere to inertial eigenstates. 
We find that inertial constraints produce efficient protocols,  which achieve the same outcome as the unoptimized protocols with reduced pulse area. Before introducing our optimization approach, we survey key aspects of the inertial theorem.

\section{The Inertial theorem}\label{sec:Inertial-thm} 

The inertial theorem exploits a temporal separation of variables to derive approximate solutions for a system's dynamics under rapid external driving~\cite{dann2021inertial}. The theorem is derived in   Liouville space -- the vector space of operators that transform states in the Hilbert space. The derivation begins by applying the Heisenberg equations of motion to the system operators, revealing a distinctive time-dependent operator basis in which the equations of motion exhibit variations on two distinct timescales.   By defining a generalized time coordinate, the rapid dynamics can be effectively removed, subsequently unveiling new dynamical symmetries—operators whose Heisenberg representation remains time-independent.  These symmetries, in turn, facilitate the construction of approximate analytical solutions.  For closed systems, we introduce a simplified formulation of the theorem in Hilbert space.  The derivation for open systems can be found in~\cite{dann2021inertial}. While our calculations account for noise, we present the Hamiltonian formulation here due to its simplicity.

Suppose $\KET{\psi}$ is a solution of  the Schr\"{o}dinger equation, $i\hbar\,\partial_t\KET{\psi} = H(t)\KET{\psi}$. Let us introduce the eigenvalue decomposition  of the Hamiltonian:
\begin{equation}
H(t) =P\Lambda P^{-1},
\end{equation}
where $P$ and $\Lambda$ are  instantaneous matrices of eigenvectors and eigenvalues of $H$  respectively. Let us use the eigenvector matrix  to define the state  $|\tilde{\psi}\rangle=P\KET{\psi}$, which satisfies
\begin{equation}
i\hbar\frac{\partial \,|\tilde{\psi}\rangle}{\partial t} = 
\left(
P^\dagger H P - i\hbar P^\dagger \frac{\partial P}{\partial t}
\right)|\tilde{\psi}\rangle\equiv \tilde{H}|\tilde{\psi}\rangle.
\label{eq:inertial-H0}
\end{equation}
[\eqref{inertial-H0} is valid for  any matrix $P$, not necessarily the eigenvector matrix.] Inertial frames are those in which $\tilde{H}$ can be written in the form 
\begin{equation}
\tilde{H} = \Omega(t) \mat{M}(\chi),
\label{eq:factor-H}
\end{equation}
where \emph{$\Omega(t)$ may vary rapidly while $\chi$ is nearly stationary}. Introducing a rescaled time coordinate, $\tau = \int_0^t \Omega(t')dt'$,  we obtain 
\begin{equation}
i\hbar\frac{\partial |\tilde{\psi}\rangle}{\partial\tau} = \mat{M}(\chi) |\tilde{\psi}\rangle.
\end{equation}
If $\mat{M}$ changes slowly compared to its energy gap, a system initialized in an eigenstate clings to it during the dynamics.  The escape probability   from eigenstate $|\tilde{\psi}_n(t)\rangle$ with eigenenergy $\tilde{\varepsilon}_n$ is given by the inertial  parameter
\begin{equation}
\eta_I^{(n)} \equiv \max_{ m \neq n} 
\left\{ \frac{\BRA{\tilde{\psi}_m}\tfrac{\partial \mat{M}}{\partial \tau}\KET{\tilde{\psi}_n}}{(\tilde{\varepsilon}_m - \tilde{\varepsilon}_n)^2} \right\},
\label{eq:inertial-param}
\end{equation}
where the ratio is maximized over all states $m\neq n$. In comparison, an eigenstate $\KET{\psi_n(t)}$ of the lab-frame Hamiltonian, $H$, with eigenenergy $\varepsilon_n$  evolves adiabatically provided that the adiabaticity parameter is small
\begin{equation}
\eta_A^{(n)} \equiv \max_{ m \neq n} 
\left\{ \frac{\BRA{\psi_m}\tfrac{\partial H}{\partial t}\KET{\psi_n}}{(\varepsilon_m-\varepsilon_n)^2} \right\}.
\label{eq:adiabatic-param}
\end{equation}
We will see that in    STIRAP,  keeping the inertial parameter small amounts to constrain the second-order temporal derivative of the control parameters (i.e.,  acceleration), while the adiabatic condition implies a small velocity.

\section{Inertial conditions for STIRAP}\label{sec:inertial-STIRAP}

Before introducing the inertial protocol, we review the essentials of STIRAP.  In this protocol, the population is transferred between two electronic states,  $\KET{1}$ and $\KET{3}$, by driving transitions from these states into a third level $\KET{2}$.  Introducing the vector  $\vec{c} = [c_1(t),c_2(t),c_3(t)]$ for  probability amplitudes of the electronic states, the system dynamics is governed by 
\begin{gather}
i\hbar\frac{d}{dt}\vec{c}(t) = \frac{\hbar}{2}
\left(\begin{array}{ccc}
    0 & \Omega_1(t) & 0 \\
    \Omega_1(t) & 2\Delta & \Omega_2(t) \\
    0 & \Omega_2(t) & 0 
\end{array}\right) \vec{c}(t).
\label{eq:STIRAP-Hamiltonian}
\end{gather}
Here, $\Omega_1$  is the Rabi frequency of the field that couples $\KET{1}$ and $\KET{2}$ while 
$\Omega_2$  couples $\KET{2}$ and $\KET{3}$. Each field is detuned from the driven atomic transition by $\hbar\Delta$, and we assume that the two-photon transition from $\KET{1}$ to $\KET{3}$ is resonant.  This Hamiltonian  has a zero-energy eigenstate $\KET{D} \equiv (\Omega_2\KET{1} - \Omega_1\KET{3})/\Omega$, denoted  ``dark'' because  it is decoupled from the radiation field.    To transition adiabatically from $\KET{1}$ to $\KET{3}$ by following the dark state, one needs to start the protocol with $\Omega_1 = 0$ and $\Omega_2 \neq 0$ (with  $\KET{D}=\KET{1}$) and end with $\Omega_2 = 0$ and $\Omega_1 \neq 0$  (so that $\KET{D}=\KET{3}$). 

Let us now construct an inertial STIRAP protocol.   For simplicity,  we present a derivation of our protocol in the limit of $\Delta = 0$, which is optimal in some cases~\cite{vitanov2017stimulated}, but we deal later with  $\Delta\neq0$ which is the relevant regime for the STIRAP-based  CZ gate (CZ)~\cite{rao2014robust}. When $\Delta=0$, the dynamics can be described by an effective two-level system, satisfying~\cite{laine1996adiabatic,vasilev2009optimum}:
\begin{gather}
i\hbar\frac{d}{dt}\vec{b}(t) = \frac{\hbar}{2}
\left(\begin{array}{cc}
     \Omega_1(t) & \Omega_2(t) \\
    \Omega_2(t) &  -\Omega_1(t)
\end{array}\right) \vec{b}(t).
\label{eq:STIRAP-H-2level}
\end{gather}
Here, $\vec{b}$ is the vector of probability amplitudes of two specially chosen basis states, expressed in terms of  $\vec{c}$ (the explicit relation is given  in~\cite{laine1996adiabatic}). While state transfer in the original basis  [\eqref{STIRAP-Hamiltonian}] requires the initial and final conditions $c(0) = [1,0,0]$ and $c(t_f) = [0,0,1]$ respectively   (where $t_f$ denotes protocol duration), the new basis coefficients  satisfy  $b_1(0) = 1$ and $b_1(t_f) = b_2(t_f)$. We  introduce the parameterization
\begin{gather}
\Omega_1 = \Omega(t)  \sin[2\theta(t)]  \quad
\Omega_2 = \Omega(t)  \cos[2\theta(t)],
\label{eq:sinusoidal}
\end{gather}
with $\theta(0) = 0$ and $\theta(t_f) = \pi/2$. 
With this parameterization, \eqref{STIRAP-H-2level} becomes
\begin{equation}
H = \Omega(t) 
\left(\cos2\theta\sigma_z + \sin2\theta\sigma_x\right).
\label{eq:2-level-lab-frame}
\end{equation}
Seeking an inertial solution, we transform the Hamiltonian to the eigenvector basis and find 
\begin{gather}
\tilde{H} = \Omega(t) \left(\sigma_z - \tfrac{\chi}{2}\sigma_y\right) = \Omega(t)\mat{M}(\chi).
\label{eq:inertial-H}
\end{gather}
We define $\chi \equiv  \tfrac{ \dot{\theta}}{\Omega}$, where dot denotes a time derivative. Derivation details are shown in~\appref{inertial-STIRAP}. When $\chi$ is small, the protocol is adiabatic. Yet, when $\chi$ varies slowly, it is 
inertial.
An inertial  protocol succeeds provided that  
(\emph{i}) the initial state is $\KET{1}$ and the final state is $\KET{2}$; (\emph{ii}) the eigenstates of $\mat{M}(\chi)$ and $H$ coincide  at the beginning and end of the protocol; and (\emph{iii}) the matrix $\mat{M}(\chi)$ varies  slowly compared to its energy gap. Formally, the conditions are:
\begin{subequations}
\begin{align}
&(i)\hspace{0.2in}\theta(0) = 0 \hspace{0.25in} ,\hspace{0.25in} \theta(t_f)= \pi/2 \\
&(ii)\hspace{0.2in}\dot{\theta}(0)=
\dot{\theta}(t_f) = 0, \label{eq:init-fin-conditions}
\\
&(iii)\hspace{0.2in}
\eta_I = \tfrac{1}{4\Omega} \tfrac{\dot{\chi}}{4+\chi^2}\ll1
\quad\mbox{where\,\,\,}
\chi \equiv  \tfrac{ \dot{\theta}}{\Omega}
\end{align}
\label{eq:inertial-conditions}
\end{subequations}
The protocol is also adiabatic when
\begin{equation}
(iv)\hspace{0.2in}\eta_A = \tfrac{ \chi}{4}\ll1\hspace{1.4in}
\end{equation}
The adiabatic and inertial parameters, $\eta_A$ and $\eta_I$ respectively, were computed using \eqsref{inertial-param}{adiabatic-param}.
In the next section, we compare the performance of several protocols. We find  that protocols with a small inertial  parameter 
have a favorable asymptotic scaling of the infidelity and are robust to realistic noise channels.

\section{Fidelity of inertial STIRAP pulses}\label{sec:benchmark}

\figrefbegin{Fig1}  compares   STIRAP performance with analytic pulse shapes and with an optimized inertial pulse. The analytic functions, $\Omega_j$ (j=1,2), are  Gaussian, squared sinusoidal (sinSQ) with a linear $\theta(t)$ and  a  sin function  with a cubic  $\theta(t)$:
 \begin{subequations}
\begin{gather}
\mathrm{Gaussian}:\hspace{0.05in}
\Omega_{j}(t) = \Omega_{\mathrm{max}} e^{-4(t -  t_j)^2/{t_f^2}} \\
\mathrm{SinSQ}:
\hspace{0.05in}
\Omega_{j}(t) = \Omega_{\mathrm{max}}\sin^2[\pi(t-t_j)/2t_f]\\
\mathrm{Cubic}:\hspace{0.05in}
\Omega_{j}(t) = \Omega_{\mathrm{max}}\sin[\theta_n(t-t_j)].
\label{eq:cubic-protocol}
\end{gather}
\label{eq:pulse-shapes}
\end{subequations}
\hspace*{-8pt} Here,  
$t_1 = t_f $ and $t_2 = 0$ while 
$\theta_n$ is a cubic polynomial. The argument of the cubic pulse is: $\theta_n(t) = \displaystyle\sum_{n=2}^3  C_nt^n$ with $ C_3 = -\tfrac{\pi}{t_f^3}, C_2 = (\tfrac{\pi}{2} - C_3 t_f^3)/t_f^2$. The time delay between Gaussian pulses is optimized to maximize the fidelity~\cite{vitanov2017stimulated}. Conditions (\emph{i}) and (\emph{ii}) are satisfied by the sinSQ and cubic pulses, but are violated by the Gaussian pulse. The optimized pulse is found with our constrained QOCT code, which minimizes the time-averaged inertiality parameter $\eta_I$, and uses the cubic pulse as an initial guess.

\begin{figure}[htbp]  \includegraphics[scale=0.58]{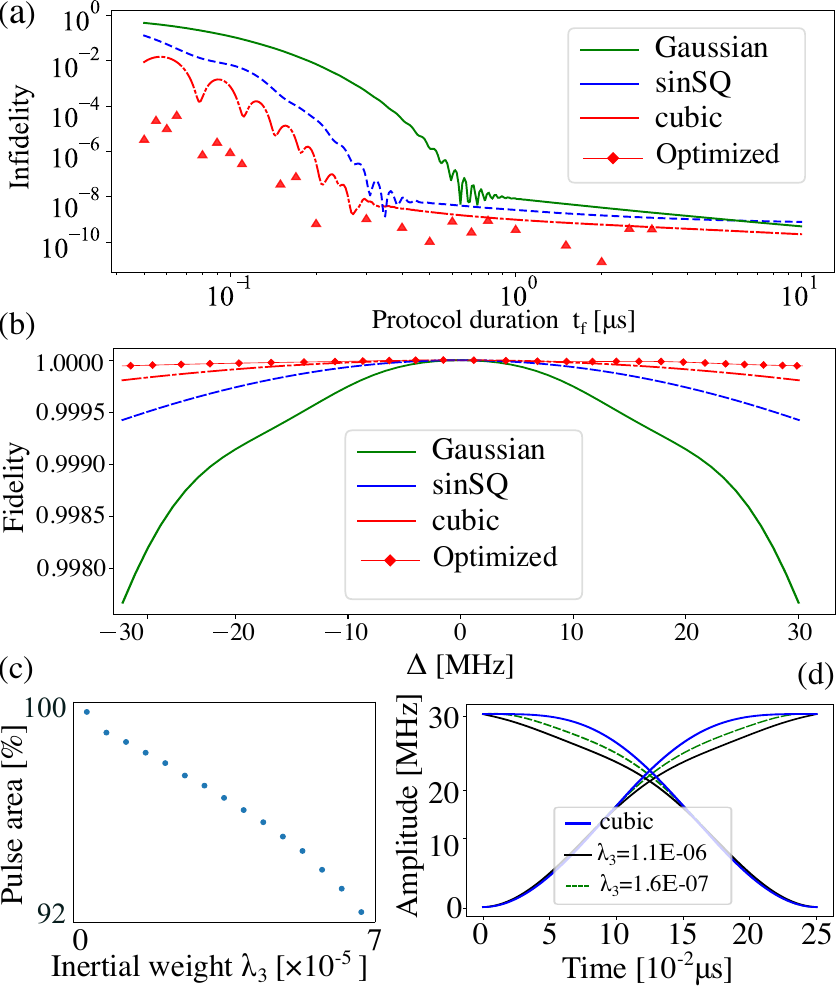}
  \caption{(a)  STIRAP infidelity versus effective pulse area ($\Omega_{\mathrm{max}}\cdot t_f$) for four pulse shapes [\eqref{pulse-shapes}]: Gaussian (green), sinSQ (blue),    cubic (red curve), and numerically optimized (red triangles).  Parameters: $\Delta = 0 \mbox{  MHz},$  $\Omega_{\mathrm{max}} = 50\mbox{  MHz}$ and $\gamma_{31}/2\pi = \gamma_{32}/2\pi = 3\, \mbox{  MHz}$, accounting for decay from $\KET{3}$ into $\KET{1}$ and $\KET{2}$.   Parameters of the optimization code:  $\lambda_1 = 0.1, \lambda_2 = 0, \lambda_3 = 1\cdot10^{-7}$ (see~\secref{QOCT-sec}). (b) Infidelity, $1-\mathcal{F}$,  versus $\Delta$, for $t_f = 0.25  \,\mu$s.
  (c) Effective pulse area, $\Omega_{\mathrm{max}}t_f$,  (relative to cubic pulse) for protocols generated with constrained QOCT vs. weight of inertial constraint,  $\lambda_3$. 
  (d) Pulse shapes generated with QOCT with a  cubic guess pulse for different values of $\lambda_3$, which achieve fidelity of $0.99$. }
   \label{fig:Fig1}
\end{figure}

To compute the infidelity, we evolve the density matrix (representing the state of the system) by solving the Lindblad master equation, with the Hamiltonian of \eqref{STIRAP-Hamiltonian} and including spontaneous emission from state $\KET{2}$.  We use the open-source Python package QuTip~\cite{johansson2012qutip}. 
We compute the infidelity using the metric
 $\mathcal{F}\equiv 1 - \mbox{Tr}{({\sqrt{\rho_f}\rho_t\sqrt{\rho_f})}^{1/2}}$
 where the target state is $\rho_t = \KET{3}\BRA{3}$ and $\rho_f$ is the simulated final state.

\figrefbegin{Fig1} compares the infidelity and robustness of the protocols as a function of protocol duration $t_f$.  In contrast to the Gaussian pulse, both the cubic and sinSQ pulses satisfy conditions (\emph{i-ii}),  and are, therefore, superior to it. The cubic pulse is better than the sinSQ pulse. This can be attributed to the fact that the energy gap between instantaneous eigenvalues is constant throughout the protocol and, hence, minimizes diabatic transitions~\cite{guerin2002optimization}. The optimized pulse has the smallest fidelity and highest robustness. Small oscillations in the infidelity of the analytic pulses are attributed to non-adiabatic transitions that can be controlled by using mask functions that smooth the derivatives of the pulses~\cite{laine1996adiabatic}.

To further substantiate the connection between inertiality and optimal adiabatic performance, we use our QOCT algorithm [\secref{QOCT-sec}]. We control the degree of inertiality by increasing the weight of the inertial constraint  [$\lambda_3$ in \eqref{inertial-QOCT} below] and find that by increasing  $\lambda_3$,  the pulse area is reduced for a given fidelity. This is the main advantage of our optimized protocols, as it implies that they are more efficient than the analytic pulse shapes. 
\figrefbegin{Fig1}(c) compares pulses obtained with varying inertiality constraints in the range $\lambda_3\in(0,7)\times10^{-5}$. 
It presents the effective pulse area, $\Omega_{\mbox{\small{max}}}t_f$, relative to that of the cubic protocol  [\eqref{cubic-protocol}], as a function of $\lambda_3$. The pulse area is reduced in the shown parameter range, as a result of increasing $\lambda_3$. (We note that this trend does not hold in general. Since we are dealing with an optimization problem with multiple objective functions, it has multiple local minima. We find  optimal weights for optimization through trial and error.) 
\figrefbegin{Fig1}(d) shows three representative optimized pulses, obtained with inertial weights of $\lambda_3 = 1.6\cdot10^{-7}$ (green dashed)  and $\lambda_3 = 1.1\cdot10^{-6}$ (black solid). We show also the cubic pulse shape (blue). While all shown pulses achieve  STIRAP fidelity of 0.99,  the smallest effective pulse area is achieved with the largest $\lambda_3$.  Our QOCT algorithm is explained in the following section.

\section{QOCT  with   inertial constraints\label{sec:QOCT-sec}}

The goal of QOCT is to find optimal pulses, $\Omega_1^c$ and $\Omega_2^c$, and a corresponding optimal trajectory, $\rho^c(t)$, that achieve maximal fidelity given minimal input power. Here, we introduce additional constraints to find optimal solutions that are also adiabatic and inertial.
Consider   the  functional
\begin{gather}
\Scale[1]{\mathcal{J}(\rho,\Omega_1,\Omega_2) = \mathrm{Tr}\{\rho_t \rho_f\} - 
\int_0^{t_f}\! dt \,\mathrm{Tr}\{
\xi(\tfrac{d}{dt}-\hat{\mathcal{L}}
)\rho\}}+\nonumber\\
\Scale[1]{
\sum\limits_{i=1}^2\left\{\lambda_1 \int_0^{t_f}\!\! dt|\Omega_i|^2+
\lambda_2 \int_0^{t_f}\!\! dt|\dot{\Omega}_i|^2+
\lambda_3 \int_0^{t_f}\!\! dt|\ddot{\Omega}_i|^2\right\}}
\label{eq:inertial-QOCT}
\end{gather}
The first term is maximized when the final state $\rho_f$  reaches the target state $\rho_t$. The second is maximized when $\rho$ satisfies the Lindblad equation, $\dot{\rho}=\hat{\mathcal{L}}\rho$, where  $\hat{\mathcal{L}}$ is the Lindbladian operator and  $\xi$  is the associated  Lagrange multiplier. The third is a penalty proportional to the total input power in the pulses, with the associated Lagrange multiplier $\lambda_1$. We include the last two terms to restrict the time-averaged velocity and acceleration, with Lagrange multipliers $\lambda_2$ and $\lambda_3$ respectively.

Out of the different methods to formulate and solve the optimization problem~\cite{khaneja2005optimal,rach2015dressing,caneva2011chopped,muller2022one}, we chose to use  Krotov’s algorithm to find minimal solutions for  $\mathcal{J}$ ~\cite{krotov1983iterative,krotov1988technique,krotov1989global,tannor1992control,krotov1995global,bartana1997laser,konnov1999global}, which starts with guess pulses for the controls and updates them iteratively. This method has the advantage that it guarantees that the performance merits improvement in each iteration~\cite{reich2012monotonically}. 
Starting from initial guess pulses, $\Omega_1^0$ and $\Omega_2^0$, in each iteration (denoted by $k$), the variables $\rho^{k}$ and $\chi^k$ are evolved and the pulses are updated according to an update rule which guarantees that $\mathcal{J}$ decreases in each iteration. Since the pulses are generally complex, we obtain separate update rules for their real and imaginary components, $\delta\Omega_i^R(t)$ and $\delta\Omega_i^I(t)$ respectively. In the following section, we derive the update rules for the fields.

\subsection*{VI.A. Derivation of the update equations}

To obtain the update rules, we compute the complete differential of $\mathcal{J}$. The update rules are obtained by equating the  partial derivatives  $\partial\mathcal{J}/\partial(\delta\Omega_i^R(t))$ and $\partial\mathcal{J}/\partial(\delta\Omega_i^I(t))$  to zero. Our derivation is a generalization of  Krotov's algorithm presented in  \cite{bartana1997laser}.  To begin, we integrate by parts \eqref{inertial-QOCT} and rewrite the first line of \eqref{inertial-QOCT} as  
\begin{gather}
\mathrm{Tr}
\left\{\rho_t \rho_f - 
(\xi\,\rho)|_0^{t_f}+
\int_0^{t_f}\!\! dt \left[
\rho\tfrac{d\xi}{dt}+
\xi\hat{\mathcal{L}}\rho\right]\right\}.
\label{eq:integration-by-parts}
\end{gather}
To derive a discretized representation for the differential, we calculate the variation in $\mathcal{J}$ from iteration $k+1$ to iteration $k$.
Using the definitions
\begin{gather}
\delta\mathcal{J}=\mathcal{J}^{(k+1)}-\mathcal{J}^{(k)}\quad,\quad
\delta\rho=\rho^{(k+1)}-\rho^{(k)},
\end{gather}
 we obtain:
To obtain the update rules, we compute the complete differential of $\mathcal{J}$. The update rules are obtained by equating the  partial derivatives  $\partial\mathcal{J}/\partial(\delta\Omega_i^R(t))$ and $\partial\mathcal{J}/\partial(\delta\Omega_i^I(t))$  to zero. Our derivation is a generalization of  Krotov's algorithm presented in  \cite{bartana1997laser}.  To begin, we integrate by parts \eqref{inertial-QOCT} and rewrite the first line of \eqref{inertial-QOCT} as  
\begin{gather}
\mathrm{Tr}
\left\{\rho_t \rho_f - 
(\xi\,\rho)|_0^{t_f}+
\int_0^{t_f}\!\! dt \left[
\rho\tfrac{d\xi}{dt}+
\xi\hat{\mathcal{L}}\rho\right]\right\}.
\label{eq:integration-by-parts}
\end{gather}
To derive a discretized representation for the differential, we calculate the variation in $\mathcal{J}$ from iteration $k+1$ to iteration $k$.
Using the definitions
\begin{gather}
\delta\mathcal{J}=\mathcal{J}^{(k+1)}-\mathcal{J}^{(k)}\quad,\quad
\delta\rho=\rho^{(k+1)}-\rho^{(k)},
\end{gather}
 we obtain:
 \begin{gather}
     \delta\mathcal{J} = \mathrm{Tr}
\left\{\rho_t \delta\rho - 
\xi^k\,\delta\rho]|_0^{t_f}\right\}\\+
\nonumber\mathrm{Tr}
\left\{\int_0^{t_f}\!\! dt \left[
\delta\rho\tfrac{d\xi}{dt}+
\xi^k\hat{\mathcal{L}}\delta\rho+
\xi^k\delta\hat{\mathcal{L}}\rho^{(k+1)}\right]\right\}\\
-\int_0^{t_f}
\left\{2\lambda_1\mathrm{Re}[^k\delta\Omega^*]+\lambda_1[\delta\Omega_R^2+\delta\Omega_I^2]\right\}\nonumber\\
-\int_0^{t_f}
\left\{2\lambda_2\mathrm{Re}[\dot{\Omega}^k\delta\dot{\Omega}^*]+\lambda_2[\delta\dot{\Omega}_R^2+\delta\dot{\Omega}_I^2]\right\}\nonumber
\\-\nonumber\int_0^{t_f}
\left\{2\lambda_3\mathrm{Re}[\ddot{\Omega}^k\delta\ddot{\Omega}^*]+\lambda_3[\delta\ddot{\Omega}_R^2+\delta\ddot{\Omega}_I^2]
\right\}.
 \end{gather}

\begin{figure*}
  \includegraphics[width=\textwidth]{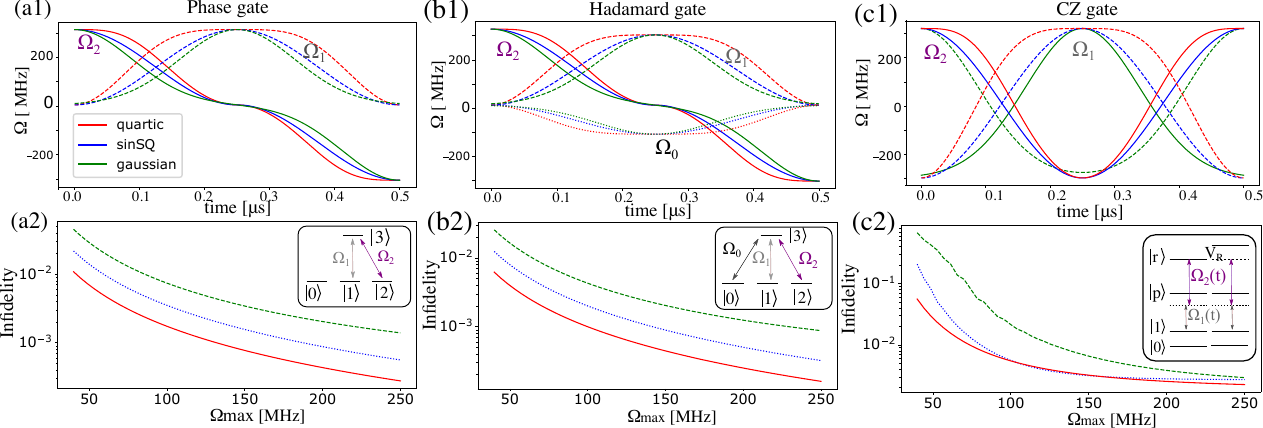}
                  \caption{
                  \textbf{Inertial geometric gates} (a1,a2) Single-qubit phase gate, using a tripod-level atom (inset) with control pulses $\Omega_1(t)$ and $\Omega_2(t)$ that transfer population from $\KET{1}$ to $\KET{2}$ and back to $\KET{1}$, while flipping the phase of $\Omega_2$ in the middle of the protocol~\cite{duan2001geometric,moller2007geometric}.
                     (a1) Gaussian (Green), sinSQ  (blue) and quartic (red) pulses are shown. (a2) Gate infidelity versus  $\Omega_{\mbox{\small{max}}}$ for the Gaussian, sinSQ, and quartic pulses. (b1-b2) Hadamard gate, using $\Omega_0$, $\Omega_1$ and $\Omega_2$  and gate infidelity. (c1-c2) CZ gate, using 4-level atoms with  Rydberg-induced phase shifts (inset). Pulse shapes and gate infidelity versus $\Omega_{\mbox{\small{max}}}$  are shown. Parameters of single-qubit gates: $\gamma/2\pi = 6$  MHz, $\Delta = 0$. Phase gate: $\Omega_1^{\mbox{\small{max}}} = \Omega_2^{\mbox{\small{max}}} = 2\pi\times50$ MHz. Hadamard gate: $\Omega_1^{\mbox{\small{max}}} = 2\pi\times50$ MHz, $\Omega_2^{\mbox{\small{max}}}=(1-\sqrt{2})\Omega_1^{\mbox{\small{max}}}$, $(\Omega_0^{\mbox{\small{max}}})^2=(\Omega_1^{\mbox{max}})^2+(\Omega_2^{\mbox{\small{max}}})^2$. Two-qubit CZ gate: $\Omega_{\mbox{\small{max}}}/2\pi = \Delta = 50$ MHz, $V_R = 4/{2\pi}$ MHz, $\gamma_p/2\pi = 6$ MHz, $\gamma_r/2\pi = 1$ kHz, $\gamma_{\mbox{\small{dep}}}/2\pi = 10$ kHz.
                  }
                  \label{fig:Fig3}
  \end{figure*}
The first four terms do not depend on the field	increment and, hence do not contribute to the update equations. We denote the remaining terms as:

\begin{gather}
\delta\mathcal{J}_1= \mathrm{Tr}
\left\{\int_0^{t_f}\!\! dt 
\xi^k\delta\hat{\mathcal{L}}\rho^{(k+1)}\right\}\nonumber
          \\  \delta\mathcal{J}_2  = -2\mathrm{Re}\left[\int_0^{t_f}
\lambda_1\Omega^k\delta\Omega^*
+\lambda_2\dot{\Omega}^k\delta\dot{\Omega}^*
+\lambda_3\ddot{\Omega}^k\delta\ddot{\Omega}^*\right]
\nonumber \\          
\delta\mathcal{J}_3= -\int_0^{t_f}\!
\lambda_1\delta{\Omega}_R^2+\!
\lambda_2\delta\dot{\Omega}_R^2+\!
\lambda_3\delta\ddot{\Omega}_R^2\ \nonumber       \\  \delta\mathcal{J}_4 = -\int_0^{t_f}\!
\lambda_1\delta{\Omega}_I^2+\!
\lambda_2\delta\dot{\Omega}_I^2+\!
\lambda_3\delta\ddot{\Omega}_I^2.   \nonumber  
\end{gather}

The surface terms vanish since the field update at the start and end point is zero, $\delta\Omega_R(0)= \delta\Omega_R(t_f) = 0$. Using this result (and repeating the same procedure for $\delta\mathcal{J}_4$), we find update equations:
\begin{gather}
-\lambda_1\delta\Omega_R+\lambda_2\delta\ddot{\Omega}_R-\lambda_3\delta\!\ddddot{\Omega}\!\!\!_R = \tfrac{1}{2}\mathrm{Tr}
\left[\xi^{k}\tfrac{\partial\delta\hat{\mathcal{L}}}{\partial\delta\Omega_R}
\rho^{k+1}\right]\equiv F(t)\nonumber\\
-\lambda_1\delta\Omega_I+\lambda_2\delta\ddot{\Omega}_I-\lambda_3\delta\!\ddddot{\Omega}\!\!\!_I = \tfrac{1}{2}\mathrm{Tr}
\left[\xi^{k}\tfrac{\partial\delta\hat{\mathcal{L}}}{\partial\delta\Omega_I}
\rho^{k+1}\right]\equiv G(t)
\end{gather}
These equations are easily solved by discretizing time and rewriting the equation in matrix form:
\begin{gather}
    \mathbf{M}_1\delta\Omega_R = \mathbf{F}\nonumber\\
    \mathbf{M}_2\delta\Omega_I = \mathbf{G}
\end{gather}
The matrices $\mathbf{M}_1$ and $\mathbf{M}_2$ are written in terms of the identity matrix, the tridiagonal second-order finite-difference derivative, and a 5-diagonal fourth-order derivative matrix.
The matrices $\mathbf{F}$ and $\mathbf{G}$ are diagonal matrices defined using $F(t)$ and $G(t)$.  
The equations are solved by inverting $\mathbf{M}_1$ and $\mathbf{M}_2$ in each iteration to find the update rules $\delta\Omega_R$ and $\delta\Omega_I$.

\section{Geometric quantum logic  gates}\label{sec:Geometric-gates} 
Robustness to noise can be achieved by  cyclic adiabatic evolution (which starts and ends with the same Hamiltonian) under which a qubit’s state acquires a geometric phase that depends only on the state's trajectory~\cite{kato1950adiabatic}. While non-degenerate Hamiltonians are associated with scalar (abelian) phases, degeneracy gives rise to matrix (non-abelian) phases~\cite{zanardi1999holonomic}. When including both abelian and non-abelian ``holonomic'' operations, it is possible to realize a universal set of  gates that are robust against certain types of errors~\cite{saffman2009efficient,rao2013dark,rao2014deterministic,petrosyan2017high,khazali2020fast}. A protocol for holonomic quantum logic gates was proposed in~\cite{moller2007geometric}, where the qubit state acquires geometric phases by driving transitions out and back into the qubit subspace. 
We simulated this protocol using the pulses from~\secref{benchmark}.

\figrefbegin{Fig3} presents numerical simulations of the gates. Single-qubit gates (panels a--b) require a  tripod-level atom (insets in the bottom plots).  A qubit is encoded in  $\KET{0}$ and $\KET{1}$. A geometric phase gate (Pauli-Z rotation by $\pi$) is performed using two STIRAP steps to transfer population from $\KET{1}$ to the auxiliary level $\KET{2}$ and back into $\KET{1}$; The phase of one of the pulses (here $\Omega_2$) is shifted by $\pi$ in the second  STIRAP step [\figref{Fig3}(a), top] and, consequently, creates a relative phase shift of $\pi$ between $\KET{0}$ and $\KET{1}$.  To find favorable performance, we replace the cubic pulse with a quatric polynomial. Since our protocol contains two STIRAP steps, we search for a polynomial with zero derivatives ($\theta'(t)=0$)  at the beginning, middle, and  end of the protocol. Considering symmetric and antisymmetric pulses (around $t_f/2$), we find that these conditions are  satisfied with a fourth order polynomial. $\theta(t)$ is a quartic polynomial $P_4(t) = C_0 + C_2(t-\tfrac{t_f}{2})^2+C_4(t-\tfrac{t_f}{2})^4$, with $C_0 = \tfrac{\pi}{2}, C_2 = -4\tfrac{\pi}{t_f^2}, C_4 = \tfrac{8\pi}{t_f^4}$.

To implement a Hadamard gate [\figref{Fig3}(b)], one needs a third field, $\Omega_0$. The three fields, $\Omega_{0,1,2}$, are applied following the protocol in (b1). 
In this case, the initial dark  state is a  combination of  $\KET{0}$ and $\KET{1}$, whose weights are determined by the  ratio of $\Omega_0$ and $\Omega_1$. During the protocol, a phase is acquired by the dark state, which   translates  into a rotation in the qubit basis \cite{moller2007geometric,toyoda2013realization}. Finally, a  CZ gate is achieved using four-level atoms with Rydberg interaction between  excited states [\figref{Fig3}(c)]. This gate operates in a regime of a weak Rydberg interaction compared to the Rabi frequencies, $V_R\ll\Omega_{1},\Omega_{2}$. In this limit, when two atoms occupy the excited  state, their interaction produces an energy shift and  the wavefunction acquires a phase proportional to  $V_R$~\cite{saffman2010quantum}.  A CZ gate is implemented by applying two STIRAP steps  and tuning the time delay between  the pulses so  that  the $\KET{11}$ component of the wavefunction acquires a $\pi$ phase shift relative to  other qubit states.

We simulated  these protocols  with the  pulses from~\eqref{pulse-shapes}.  For each protocol, we reconstructed the simulated gate, $\rho\rightarrow \hat{G}(\rho)$, with process tomography (reviewed in \appref{tomography}), which  provides  Kraus operators, $G_k$ that expand   $\hat{\mathcal{G}}$:~\cite{nielsen2002quantum}
\begin{equation}
\hat{\mathcal{G}}(\rho) = \sum_k G_k\rho G_k^\dagger.
\end{equation}
Using this expansion, we computed the average gate fidelity, $\mathcal{F}$, defined as the overlap between the simulated and target gates ($\hat{\mathcal{G}}$ and $U_0$ respectively) averaged   over all possible initial states. For an initial pure state, $\KET{\psi}$, the average gate fidelity $\mathcal{F}$ is given by~\cite{nielsen2002simple,beterov2016simulated,goerz2014robustness,pedersen2007fidelity}:
\begin{gather}
\mathcal{F} \equiv 
\int_{S^{2n-1}} \hspace{-0.2in}dV\,
\BRA{\psi}
U_0^\dagger
\hat{G}(\KET{\psi}\!\BRA{\psi})
U_0\KET{\psi}
=
\nonumber\\
\frac{1}{n(n+1)}\sum_k\left(
\mbox{Tr}(M_kM_k^\dagger)+\left|\mbox{Tr}(M_k)\right|^2
\right),
\end{gather}
where $n$ is the dimension of the Hilbert space ($n = 2$ or 4 for  single-   or  two-qubit gates) and   $M_k \equiv U_0^\dagger G_k$.  The results are shown in \figref{Fig3}.  We plot the infidelity as a function of maximum Rabi frequency. In the CZ gate,    $t_f$ is set by requiring that the conditional phase shift  is $\pi$. 
The quartic  protocol approaches $\mathcal{F}=1$ at the fastest rate, as expected based on \figref{Fig1}. Unfortunately, the gate fidelity is much smaller than the  STIRAP fidelity.
The reason is that  gate fidelities are limited by phase errors, which are  larger than population errors that limit STIRAP. It  is a consequence of first-order perturbation theory that population errors scale quadratically with  the small parameter while phase errors scale linearly.

\section{Noise analysis}\label{sec:Noise} 

Next, we   analyze the effect of noise on the CZ gate, with the pulses from~\figref{Fig3}(c). We consider qubits encoded in   hyperfine states of  $^{87}$Rb atoms, trapped in optical tweezers. Following the noise analysis from Ref.~\cite{saffman2005analysis,li2022single}, we include the following imperfections: Doppler shifts, errors due to nonuniformity of Gaussian control beams,  fluctuations in the amplitude and phase of the control fields, deviations in atom separation, dephasing and finite lifetimes of the electronic states.   Details about our noise model (which follows Ref.~\cite{li2022single})  are given in~\appref{noise}.

\begin{figure}[t]
  \includegraphics[width=0.5\textwidth]{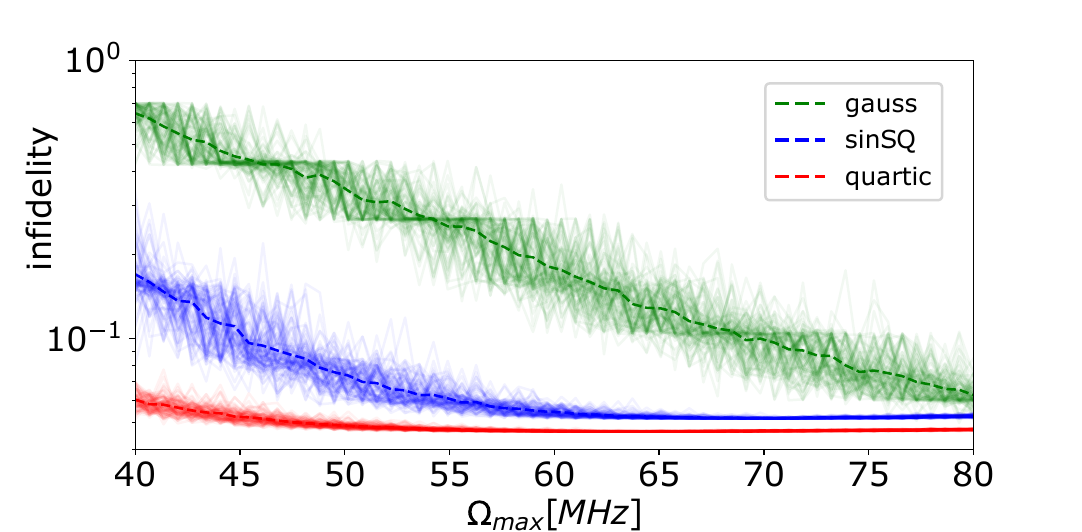}
                  \caption{Infidelity of  the CZ gate protocol [\figref{Fig3}(c)] performed with Gaussian, sinSQ, and quartic pulses, for 100 realizations of noisy model parameters. The detuning, maximum Rabi frequency, and deviation in atomic positions along and perpendicular to the  tweezers'  direction  are Gaussian random variables, with average values listed in~\tableref{Noise-table} and standard deviations:   $\sigma_{\Delta} = 14$ kHz, $\sigma_{\Omega} = 2.5$ MHz,  $\sigma_{dz} = 0.2  \,\mu$m and $\sigma_{dx} = \sigma_{dy} = 0.07  \,\mu$m respectively~\cite{levine2022dispersive,li2022single}. Realization are shown by a thin lines and their average  is shown by the thick dashed lines. }
                  \label{fig:Fig2}
  \end{figure}

The results  are shown in \figref{Fig2} and in \tableref{Noise-table}.
Evidently, the quartic protocol is highly robust to deviations in the detuning (Doppler shifts). When varying the detuning by  20$\%$, the fidelity of the quartic protocol varies by 0.08$\%$ while that of the Gaussian pulse varies by $\sim0.7\%$. In addition, the quartic pulse is highly robust to variations in intensity and  decay rate of the intermediate STIRAP level, $\KET{p}$. 
Being based on STIRAP, the population in the $\KET{p}$ state remains small during the protocol, and the decay rate from  $\KET{p}$  contributes only marginally to the error in our scheme. We assume, for simplicity, that the control fields are co-propagating to avoid fluctuations in the two-photon detuning. Counter-propagating beams were treated in~\cite{vitanov2017stimulated,li2022single}).

However, being based on the non-blockaded Rydberg  regime ($V_R<\Omega_1, \Omega_2$), our protocol is more sensitive   to  dephasing of the Rydberg state and to fluctuations in  atom separation compared to protocols that operate in the blockaded regime~\cite{saffman2005analysis,li2022single}. The latter  modify the strength of atomic interaction and, consequently, the phase acquired by the $\KET{11}$ component of the wavefucntion.   This drawback  is a property of a non-blockaded gate, and   not of the  quartic or Gaussian pulse shapes that implement it. An application of inertial conditions to optimize adiabatic protocols which operate in the blockaded regime (as in~\cite{li2016shortcut}) are expected to  be highly robust and efficient. \vspace{5pt}



\begin{table}
    \caption{Robustness of quartic and Gaussian CZ gates.}
    \begin{tabular}{ |p{1.5cm}||p{1.3cm}|p{1.3cm}||p{1.3cm}|p{1.3cm}| }
        \hline
        Pulse &  \multicolumn{2}{|c|}{Quartic} &  \multicolumn{2}{|c|}{Gaussian} \\
        \hline\hline
        noise & $\mathcal{F}_{\text{\tiny min}}$ & $\mathcal{F}_{\text{\tiny max}}$ & $\mathcal{F}_{\text{\tiny min}}$ & $\mathcal{F}_{\text{\tiny max}}$ \\
        \hline\hline
        detuning $\Delta \pm 20\%$ & \quad\quad$-0.09\%$ & \quad\quad$+0.084\%$ & \quad\quad$-0.839\%$& \quad\quad$0.64\%$\\ \hline
        intensity $\Omega \pm 20\%$ & \quad\quad$-0.19\%$ & \quad\quad$+0.11\%$ & \quad\quad$-1.68\%$& \quad\quad$0.8\%$\\ \hline   
        lifetime $\gamma_p \pm 20\%$ & \quad\quad$-0.017\%$ & \quad\quad$+0.013\%$ & \quad\quad$-0.13\%$& \quad\quad$+0.13\%$\\ \hline
        lifetime $\gamma_r \pm 20\%$ & \quad\quad$-0.025\%$ & \quad\quad$+0.028\%$ & \quad\quad$-0.025\%$& \quad\quad$+0.025\%$\\ \hline
        dephasing $\gamma_d \pm 20\%$ & \quad\quad$-0.95\%$ & \quad\quad$+0.64\%$ & \quad\quad$-0.59\%$& \quad\quad$+0.60\%$\\ \hline
        position $\Delta x \pm 2\%$ & \quad\quad$-5.2\%$ & \quad\quad$0.79\%$ & \quad\quad$-6.71\%$& \quad\quad$+0.86\%$\\ \hline
    \end{tabular}
\end{table}

When using parameters: $\Omega_{\mathrm{max}} = \Delta = 100\times2\pi$ MHz, $t_f = 0.5  \,\mu$s, $V_R = C_6/r^6$, with $C_6 = 14$ THz/$  \,\mu$m, $r = 11$ $  \,\mu$m, $\gamma_p = 6$ MHz, $\gamma_r = 1$ kHz, $\gamma_d = 10$ kHz,  the quartic pulse fidelity is  $\mathcal{F} = 0.963$ while the Gaussian pulse achieves $\mathcal{F} = 0.948$.
We use realistic parameters of $^{87}$Rb atoms with $\KET{p} = \KET{5P_{3/2}}$ and $\KET{r}$  with principle number $n = 97$~\cite{saffman2005analysis,rao2014robust}. The table shows by what percentage the fidelity changes upon changing the model parameters under realistic noise sources. 
\label{tab:Noise-table}

\section{
Realization of single-qubit inertial gates}\label{sec:realization} 

\begin{figure}[t] \includegraphics[width=0.45\textwidth]{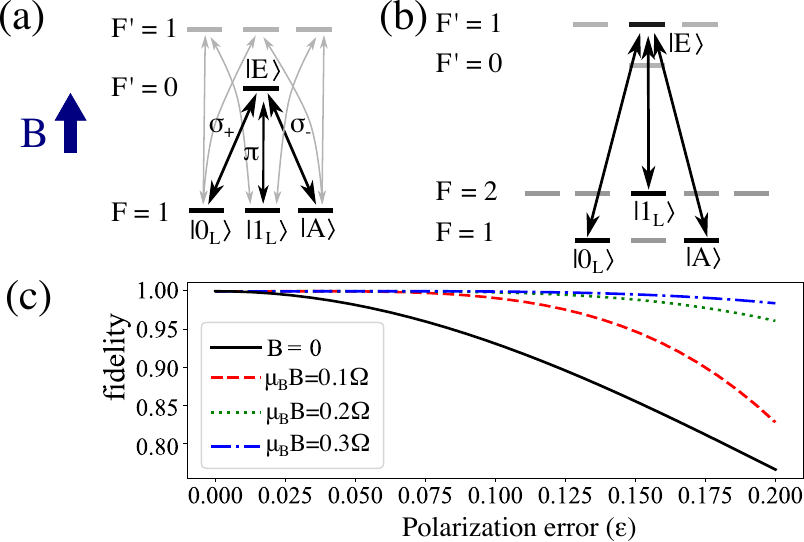}
 \caption{\textbf{Single-qubit inertial gates with Rb atoms.}} (a,b) Two possible realizations of a tripod-level system using the D2 line of  $^{87}$Rb.  Desired transitions (solid lines) are shown with the corresponding polarizations: circular ($\sigma_\pm$) and $\pi$. Additional transitions are shown by grey arrows. 
   (c) Infidelity of the phase gate [\figref{Fig2}(a)] versus polarization error, $\varepsilon$, defined as the fraction of undesired polarization in each STIRAP pulse. Nonlinear (Paschen-Back) corrections are not included in our model (see text).            
   \label{fig:Fig4}
  \end{figure}

Inertial protocols for population transfer in two-level systems were validated in recent work~\cite{hu2021experimental}. Here, We analyze a realization  of our inertial protocol for quantum logic gates. We use $^{87}$Rb atoms, with two possible choices  of energy levels and transitions from the D2 line, as shown  in \figref{Fig4}(a-b).   
In particular, we are  interested in analyzing realistic imperfections, including polarization  and  leakage errors (following the analysis from~\cite{shomroni2014all,rosenblum2016extraction,rosenblum2017analysis,bechler2018passive}). 
For example, consider a single-qubit phase gate implemented using the levels shown in  \figref{Fig4}(a). Ideally, to drive STIRAP transitions from the state $\KET{0_L}$ to $\KET{A}$ and back, we want to apply only two beams: The pulse $\Omega_1$ with  $\sigma_+$ polarization, which drives $\KET{F = 1,m_F = -1}\leftrightarrow\KET{F'=0,m_F = 0}$, and the pulse $\Omega_2$ with  $\sigma_-$ polarization,  which drives  $\KET{F = 1,m_F = 1}\leftrightarrow\KET{F'=0,m_F = 0}$.
However, polarization errors in the second pulse, $\Omega_2$, produce a $\sigma_+$ component that drives the first transition, and  errors in the first pulse drive the second transition. This process reduces the success probability of the protocol. To combat this issue, one can apply  a DC magnetic field to shift the energy levels and tune the frequencies of the control fields to be in resonance with the shifted levels. By off-setting the frequencies of the $\Omega_1$ and $\Omega_2$ pulses, the error fields become off resonant and their effect is suppressed.  
Our calculation assumes that the magnetic field is in the weak Zeeman regime.  A more accurate treatment should include nonlinear (Paschen-Back) corrections for fields exceeding   10G. Field stength of 10G corresponds to the green curve at Rabi frequency $\Omega = 50$ MHz. Our simulations demonstrate that the fidelity approaches unity when  increasing the magnetic field [\figref{Fig4}(c)].  Further improvement can be gained by introducing  a resonant  optical  cavity to enhance  desired transitions, e.g., using the setup of~\cite{bechler2018passive}. 
 The phase control suggested here is well within the capabilities of existing control.
Finally, we mention that 
at high Rabi frequencies, 
coupling between the desired Rydberg level and additional levels can cause leakage and dephasing. This effect can be treated following~\cite{walker2008consequences,vsibalic2017arc}
but is beyond the scope of the present analysis.

\vspace{0.5cm}
\section{Conclusion}\label{sec:Conclusion}

In this work, we presented  optimal adiabatic and inertial protocols for STIRAP and geometric single- and two-qubit  gates. In inertial protocols,   the system adheres to an instantaneous eigenstate of the inertial Hamiltonian, obtained from the original one by a basis change. Since  inertial protocols are not required to be adiabatic  in the original basis,  they  provide a way to construct fast high-fidelity operations.  To assess their performance, we computed  the fidelity and robustness of our protocols under various perturbations, including  fluctuations in  atomic positions and velocities,  inhomogeniety and classical noise in  the control fields, polarization impurity, spontaneous emission, dephasing, and populating leakage.   We found optimal inertial pulses with QOCT, extended to include adiabatic and inertial constraints, and presented several optimal pulses.  Our algorithm  can be further improved by searching for  solutions that maximize    the overlap between the dynamical and instantaneous inertial  eigenstates, generalizing Ref.~\cite{brif2014exploring}.

Although we find clear advantages to inertial geometric gates, some limitations of our approach should be mentioned.  The  condition that  initial and final  eigenstates of the inertial and original Hamiltonians  must coincide [\eqref{init-fin-conditions}] limits our method. Without it, inertial protocols could potentially  achieve  high fidelities very rapidly. However,  for STIRAP, this condition implies that the adiabaticity parameter must vanish at the start and end of the protocol;  Consequently,  our inertial STIRAP  protocols   are also adiabatic. To design inertial protocols that are not adiabatic, one should find   quantum algorithms  that do not start and end in  eigenstates of the lab-frame Hamiltonian. We also mention some   limitations of the geometric  protocols that were implemented~\cite{duan2001geometric,moller2007geometric,rao2014robust}.   First, our CZ gate is sensitive to errors in the time spent in the doubly excited Rydberg state.  Secondly,  two-qubit gates that  use Rydberg interactions in the dispersive regime    are sensitive to fluctuations in atomic positions. Application of  constrained  QOCT   to protocols that use Rydberg blockade is expected to  find pulses which are  robust to such fluctuations~\cite{li2022single}. Generally, inertial pulses are likely to improve many adiabatic protocols, including   topological adiabatic algorithms~\cite{kitaev2010topological,lahtinen2017short}. 

\hspace{10pt}In light of the long list of existing ``improved adiabatic protocols,''  our work adds  simple analytic pulse shapes that yield nearly optimal solutions by satisfying the   condition of small acceleration~\cite{dann2021inertial}. Our optimization algorithm, which uses the Lindblad formulation for the dynamics, finds efficient   solutions (with reduced power requirement) that correctly balance  the different  noise channels with their respective branching ratios. Given the  immense   progress in pulse-shaping techniques  for optimal control~\cite{barredo2016atom,bernien2017probing,henriet2020quantum},  our protocols are feasible   and useful for improving    the  performance merits of  quantum logic gates with atomic qubits.
Our work introduces a computational framework specifically designed to address a crucial challenge in the field – finding optimal pulses for efficient entangling gates. We believe that the tools developed in this work will provide practical solutions to complex problems. To conclude, we believe that inertial-inspired optimization tools have the potential to enhance general adiabatic protocols, thereby contributing to the advancement of state-of-the-art solutions for diverse computational challenges.

 \section*{Acknowledgement}
 The authors would like to thank Aviv Aroch for helping us set up and run Krotov's algorithm. 
 A.P. acknowledges support from the Alon Fellowship of the Israeli Council of Higher Education. 
 B.D. acknowledges support from the Israeli Science Foundation, the Binational Science Foundation, H2020
Excellent Science (DAALI, 899275), the Minerva Foundation.
 R.K. acknowledges support from the Israeli Science Foundation, grant  number 526/21.
  R.D. acknowledges support from the
 Adams Fellowship Program of the Israel Academy of Science and Humanities.

\appendix

\title{\Large \Scale[1.2]{\mbox{Supplementary Information}}}

\section{Inertial STIRAP Hamiltonian}\label{app:inertial-STIRAP}


\renewcommand{\theequation}{A\arabic{equation}}
\setcounter{equation}{0}

In this section, we  derive~\eqref{inertial-H} from the main text.  We choose the parameterization
\begin{equation}
\Omega_1 = \Omega\sin\theta \quad,\quad
\Omega_2 = \Omega\cos\theta.
\end{equation}
The 2-level Hamiltonian [\eqref{2-level-lab-frame} from the main text] becomes
\begin{equation}
H = \tfrac{\Omega^2}{2\Delta}
\left(
\cos2\theta\sigma_z + \sin2\theta\sigma_x\right)
\end{equation}
where $\sigma_x$ and $\sigma_z$ are Pauli matrices. 
We obtain an explicit expression for the inertial-frame Hamiltonian defined as 
\begin{equation}
\tilde{H}\equiv
P^\dagger H P - i\hbar P^\dagger \frac{\partial P}{\partial t}
\end{equation}
The eigenvalues are $\pm\Omega$. The eigenvector matrix is
\begin{equation}
P = \left( \begin{array}{cc}
\frac{\cot2\theta+\csc2\theta}{\sqrt{\cot{\theta}^2+1}} & \frac{\cot2\theta-\csc2\theta}{\sqrt{\tan{\theta}^2+1}}  \\
\frac{1}{\sqrt{\cot{\theta}^2+1}} & \frac{1}{\sqrt{\tan{\theta}^2+1}} \end{array} \right).
\end{equation}
For $0<\theta<\pi$, we find
\[
P^\dagger\frac{dP}{d\theta} = 
\tfrac{\dot{\theta}}{2}
\left( \begin{array}{cc}
0 & -1  \\
1 & 0 \end{array} \right).
\]
This completes the proof of \eqref{inertial-H} in the main text.

One finds a striking similarity between the inertial-frame Hamiltonian [\eqref{inertial-H}] and the Hamiltonian one obtains when adding counter-diabatic terms to the lab-frame Hamiltonian realizing the shortcut-to-adiabaticity protocol (see Eqs.~(4-5) in the Methods section of~\cite{du2016experimental}). 
The similarity is  expected, since the inner product of the instantaneous eigenvectors and their derivatives appear in both formulations. 
However, there is a conceptual difference which is important to point out. 
In shortcut-to-adiabaticity, one constructs a Hamiltonian, which may be challenging to realize, but whose eigenstates are exact solutions to the dynamics. In contrast, in our approach, the lab-frame Hamiltonian is straightforward, and the instantaneous eigenstates of the inertial-frame solutions are only approximations to the true dynamics.

\section{Adiabatic and inertial constraints in QOCT}\label{app:QOCT}
\renewcommand{\theequation}{B\arabic{equation}}
\setcounter{equation}{0}

In this section, we generalize Krotov's QOCT  algorithm  to include inertial and adiabatic constrains.  Let us consider the following functional to be minimized:
\begin{gather}
\mathcal{J} = \mathrm{Tr}\{\rho_t \rho_f\} - 
\int_0^{t_f}\!\! dt \mathrm{Tr}\{
\xi(t)\left[\tfrac{d}{dt}-\hat{\mathcal{L}}
\right]\rho(t)\}\nonumber\\
+\lambda_1 \int_0^{t_f}\!\! dt|\Omega(t)|^2+
\lambda_2 \int_0^{t_f}\!\! dt|\dot{\Omega}(t)|^2+
\lambda_3 \int_0^{t_f}\!\! dt|\ddot{\Omega}(t)|^2
\label{eq:J-appendix}
\end{gather}
where the two last terms correspond to the integrated velocity and acceleration. Using integration by parts, one can rewrite  the first line of \eqref{J-appendix} as:
\begin{gather}
\mathrm{Tr}
\left\{\rho_t \rho_f - 
(\xi\,\rho)|_0^{t_f}+
\int_0^{t_f}\!\! dt \left[
\rho\tfrac{d\xi}{dt}+
\xi\hat{\mathcal{L}}\rho\right]\right\}.
\label{eq:integration-by-parts}
\end{gather}
Krotov's algorithm starts with an initial guess from the control fields, evolves $\rho$ and $\xi$ and finds an updated control pulse that decreases $\mathcal{J}$. By iteratively updating the control pulse, convergence is reached.   Let us  compute the change in $\mathcal{J}$  between   iteration $k$ and $k+1$ in terms of the change in $\rho$. Using the definitions
\begin{gather}
\Delta\mathcal{J}=\mathcal{J}^{(k+1)}-\mathcal{J}^{(k)}\quad,\quad
\Delta\rho=\rho^{(k+1)}-\rho^{(k)},
\end{gather}
 we obtain:
 \begin{gather}
\Delta\mathcal{J} = \mathrm{Tr}
\left\{\rho_t \Delta\rho - 
\xi^k\,\Delta\rho]|_0^{t_f}\right\}+\nonumber\\
\mathrm{Tr}
\left\{\int_0^{t_f}\!\! dt \left[
\Delta\rho\tfrac{d\xi}{dt}+
\xi^k\hat{\mathcal{L}}\Delta\rho+
\xi^k\Delta\hat{\mathcal{L}}\rho^{(k+1)}\right]\right\}+\nonumber\\
-\int_0^{t_f}
\left\{2\lambda_1\mathrm{Re}[\varepsilon^k\Delta\varepsilon^*]+\lambda_1[\Delta\varepsilon_R^2+\Delta\varepsilon_I^2]\right\}+\nonumber\\
-\int_0^{t_f}
\left\{2\lambda_2\mathrm{Re}[\dot{\varepsilon}^k\Delta\dot{\varepsilon}^*]+\lambda_2[\Delta\dot{\varepsilon}_R^2+\Delta\dot{\varepsilon}_I^2]\right\}+\nonumber\\
-\int_0^{t_f}
\left\{2\lambda_3\mathrm{Re}[\ddot{\varepsilon}^k\Delta\ddot{\varepsilon}^*]+\lambda_3[\Delta\ddot{\varepsilon}_R^2+\Delta\ddot{\varepsilon}_I^2]
\right\}.
\end{gather}
The first 4 terms do not depend on the field	increment and produce the dynamical equations for $\rho$ and $\xi$. The additional terms produce the differential: $\Delta\mathcal{J}_1+
\Delta\mathcal{J}_2+\Delta\mathcal{J}_3+\Delta\mathcal{J}_4$ where
\begin{gather}
\Delta\mathcal{J}_1 = 
\mathrm{Tr}
\left\{\int_0^{t_f}\!\! dt 
\xi^k\Delta\hat{\mathcal{L}}\rho^{(k+1)}\right\}
\nonumber\\
\Delta\mathcal{J}_2 =  
-2\mathrm{Re}\left[\int_0^{t_f}
\lambda_1\varepsilon^k\Delta\varepsilon^*
+\lambda_2\dot{\varepsilon}^k\Delta\dot{\varepsilon}^*
+\lambda_3\ddot{\varepsilon}^k\Delta\ddot{\varepsilon}^*\right]\nonumber\\
\Delta\mathcal{J}_3 =  
-\int_0^{t_f}\!
\lambda_1\Delta{\varepsilon}_R^2+\!
\lambda_2\Delta\dot{\varepsilon}_R^2+\!
\lambda_3\Delta\ddot{\varepsilon}_R^2\nonumber\\
\Delta\mathcal{J}_4 =  
-\int_0^{t_f}\!
\lambda_1\Delta{\varepsilon}_I^2+\!
\lambda_2\Delta\dot{\varepsilon}_I^2+\!
\lambda_3\Delta\ddot{\varepsilon}_I^2.
\end{gather}
The term $\Delta\mathcal{J}_2$ vanishes on average since the integrand is rapidly oscillating. The complete differential of the third term is
\begin{gather}
d(\Delta\mathcal{J}_3) = \int_0^{t_f} df(\Delta\varepsilon_R,\Delta\dot{\varepsilon}_R,\Delta\ddot{\varepsilon}_R)dt = \nonumber\\
\int_0^{t_f}
\left[\tfrac{df}{d\Delta\varepsilon_R}\Delta\varepsilon_R+
\tfrac{df}{d\Delta\dot{\varepsilon}_R}\Delta\dot{\varepsilon}_R+
\tfrac{df}{d\Delta\ddot{\varepsilon}_R}\Delta\ddot{\varepsilon}_R\right]dt \nonumber\\
\int_0^{t_f}
\left[
\tfrac{\partial f}{\partial\Delta\varepsilon_R} -
\tfrac{d}{dt}\tfrac{\partial f}{\partial\Delta\dot{\varepsilon}_R} + 
\tfrac{d^2}{dt^2}\tfrac{\partial f}{\partial\Delta\ddot{\varepsilon}_R}
\right]\Delta\varepsilon_R\,dt +\mbox{surface terms}\nonumber\\
2\int_0^{t_f}(-\lambda_1\Delta\varepsilon_R+\lambda_2\Delta\ddot{\varepsilon}_R-\lambda_3\Delta\!\ddddot{\varepsilon}\!\!\!_R)+\mbox{surface terms}
\end{gather}
The surface terms vanish since the field update at the start and end point is zero, $\Delta\varepsilon_R(0)= \Delta\varepsilon_R(t_f) = 0$. Using this result (and repeating the same procedure for $\Delta\mathcal{J}_4$), we find update equations:
\begin{gather}
-\lambda_1\Delta\varepsilon_R+\lambda_2\Delta\ddot{\varepsilon}_R-\lambda_3\Delta\!\ddddot{\varepsilon}\!\!\!_R = \tfrac{1}{2}\mbox{Tr}
[\xi^{k}\tfrac{\partial\Delta\hat{\mathcal{L}}}{\partial\Delta\varepsilon_R}
\rho^{k+1}]\equiv F(t)\nonumber\\
-\lambda_1\Delta\varepsilon_I+\lambda_2\Delta\ddot{\varepsilon}_I-\lambda_3\Delta\!\ddddot{\varepsilon}\!\!\!_I = \tfrac{1}{2}\mbox{Tr}[
\xi^{k}\tfrac{\partial\Delta\hat{\mathcal{L}}}{\partial\Delta\varepsilon_I}
\rho^{k+1}]\equiv G(t)
\end{gather}
These equations are easily solved by discretizing  time and rewriting the equation in matrix form (the second-order time derivatives becomes the tridiagonal Hessian matrix and the fourth-order time derivatives is a 5-diagonal matrix).
We obtain equations of the form
\begin{gather}    \mat{M}_1\Delta\varepsilon_R = \mat{F}\nonumber\\
    \mat{M}_2\Delta\varepsilon_I = \mat{G}
\end{gather}
The update rules for the control fields are found by inverting $\mat{M}_1$ and $\mat{M}_2$.

\begin{figure*}[t]
\includegraphics[width=\textwidth]{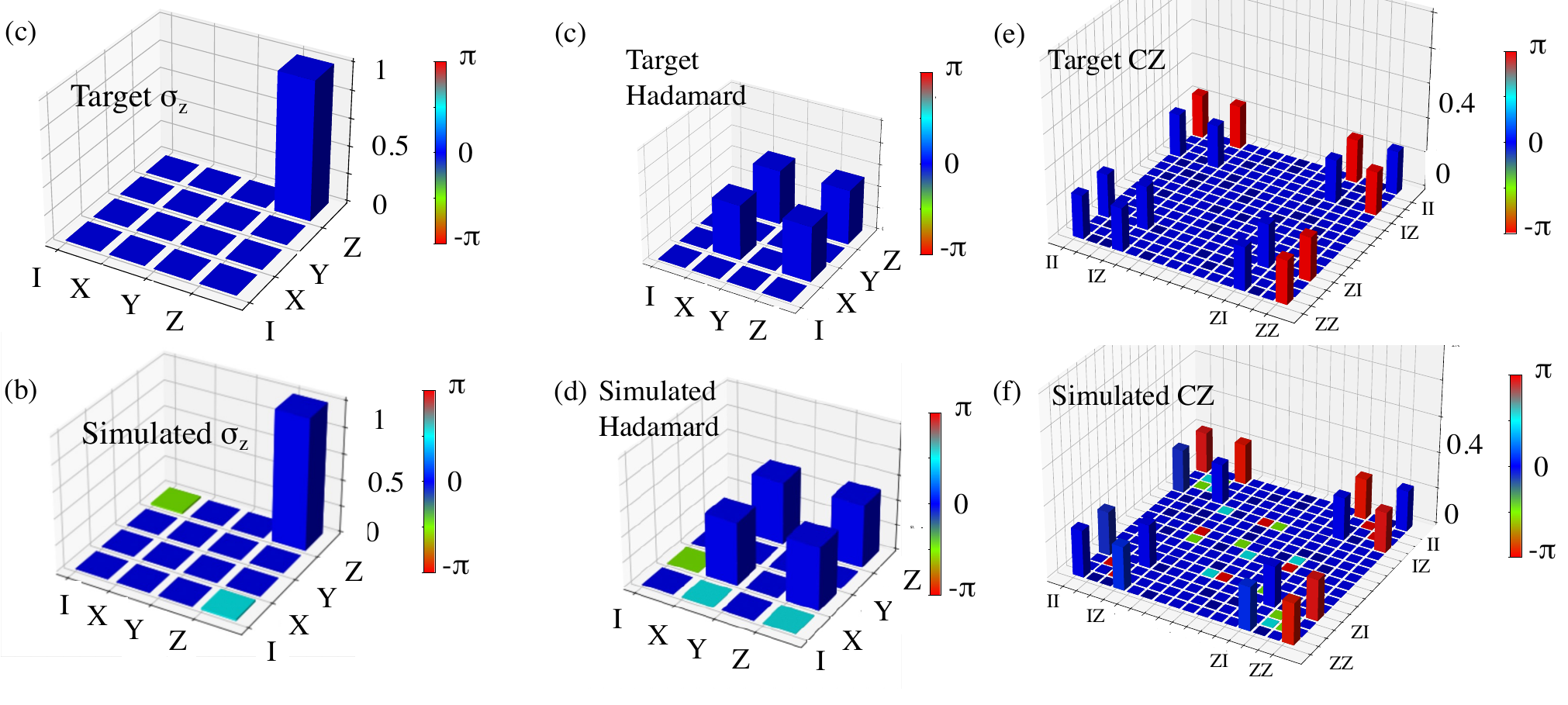}
  \caption{\textbf{Quantum gate tomography:}
  Visualization of  target (top) and simulated (bottom)  single- and  two-qubit  gates. 
The plots show the matrix $\chi$, which   expresses the   expansion coefficients of the  gate transformations in terms of Pauli basis operators,
$\tilde{E}_i$ [\eqref{Pauli-basis}] for single-qubit gates  and Kronecker products of $\tilde{E}_i$ for two-qubit gates respectively.
  (a--d) The matrix  $\chi$, 
  calculated using  \eqref{chi-single-qubit},  for Hadamard and Pauli-Z single-qubit gates. (e,f) The matrix  $\chi$, calculated via \eqref{chi-two-qubit} for a two-qubit CZ gate. }
   \label{fig:tomography}
\end{figure*}

\section{Quantum gate tomography}\label{app:tomography}
\renewcommand{\theequation}{C\arabic{equation}}
\setcounter{equation}{0}

Quantum process tomography (QPT) is a method by which a quantum gate is reconstructed from simulated data (experimental or numerical).  In this work, we use QPT to quantify the overlap between our simulated and target gates. In this appendix, we review the method presented in \cite{nielsen2002quantum}.

Suppose that  a density matrix $\rho$ evolves under our simulation  into $\mathcal{G}(\rho)$. The goal of QPT is to determine  $\mathcal{G}$.  
Given a  complete  basis of matrices, $\tilde{E}_m$,  that spans the operator  space (i.e., the Liouville space),  any unitary $\mathcal{G}$ can be written in the form
\begin{equation}
 \mathcal{G}(\rho)=\sum_{m,n} \chi_{mn}\tilde{E}_m\rho \tilde{E}_n^\dagger,
\end{equation}
where $\chi_{mn}$ are coefficients that we wish to find.  For single-qubit gates,  the set $\tilde{E_m}$ contains four  $2\times2$ matrices ($m=1,...,4$) and $\chi_{mn}$ is a $4\times4$ matrix. Let us choose the Pauli and identity matrices as basis operators:
\begin{gather}
\tilde{E}_1 = I\quad,\quad
\tilde{E}_2 = \sigma_x\quad,\quad
\tilde{E}_3 = -i\sigma_y\quad,\quad
\tilde{E}_4 = \sigma_z.
\label{eq:Pauli-basis}
\end{gather}
To determine $\chi$, we run our simulations on 4 initial states 
\begin{gather}
\rho_1 = \KET{0}\!\BRA{0}\hspace{0.05in},\hspace{0.05in}
\rho_2 = \rho_1\sigma_x \hspace{0.05in},\hspace{0.05in}
\rho_3 = \sigma_x\rho_1\hspace{0.05in},\hspace{0.05in}
\rho_4 = \sigma_x\rho_1\sigma_x.
\end{gather}
We compute the final state for every  input: $\rho_i'\equiv\mathcal{E}(\rho_i)$. Then, 
following \cite{nielsen2002quantum}, $\chi$ is given by   
\begin{equation}
\chi = 
\frac{1}{4} \begin{pmatrix}
I & \sigma_x\\
\sigma_x & I
 \end{pmatrix}
 \begin{pmatrix}
\rho_1' & \rho_2' \\
\rho_3' & \rho_4'
 \end{pmatrix}
 \begin{pmatrix}
I & \sigma_x\\
\sigma_x & I
 \end{pmatrix}.
 \label{eq:chi-single-qubit}
\end{equation}
where $I, \sigma_x, \rho'_{1},...,\rho'_{4} \in \mathbb{R}^{2\times2}$  and $\chi \in \mathbb{R}^{4\times4}$.

For two-qubit gates, 
we write
\begin{equation}
 \mathcal{G}(\rho)=\sum_{m,n} \chi_{mn}\tilde{E}_m^{(2)}\rho (\tilde{E}_n^{(2)})^\dagger
\end{equation}
in terms of two-qubit operators, $\tilde{E}_m^{(2)}\in\mathbb{R}^{4\times4}$, defined as 
Kronecker products of the single-qubit  basis operators, $\tilde{E}_m$, from \eqref{Pauli-basis}.
 The matrix $\chi\in\mathbb{R}^{16\times16}$ is given by
\begin{equation}
\chi=\Lambda\,\Bar{\rho}\,
\Lambda, 
 \label{eq:chi-two-qubit}
\end{equation}
where $\Lambda\in\mathbb{R}^{16\times16}$ is a rotation matrix
\begin{align}
\Lambda\equiv
\frac{1}{4}\left[
\begin{pmatrix}
I & \sigma_x\\
\sigma_x & I
\end{pmatrix}\otimes\begin{pmatrix}
I & \sigma_x\\
\sigma_x & I
 \end{pmatrix}\right]
\end{align}
and $\Bar{\rho}\in\mathbb{R}^{16\times16}$ is defined as
\begin{align}
\Bar{\rho}=P^T \rho' P. 
\end{align}
Here, $\rho'\in\mathbb{R}^{16\times16}$
the matrix of final states 
\begin{align}
\rho' = \mathcal{G}(\rho),
\end{align}
found by propagating  16 input  states 
\begin{align}
\rho_{mn} = T_n\KET{00}\BRA{00}T_m\quad
\forall n,m =1,...,4
\end{align}
with 
\begin{gather}
T_1=I\otimes I \quad,\quad
T_2=I\otimes\sigma_x\nonumber\\
T_3=\sigma_x\otimes I  \quad,\quad
T_4=\sigma_x\otimes \sigma_x.
\end{gather}
The permutation matrix $P$ is given by 
\begin{align}
P = I\otimes\ 
 \left[  \begin{pmatrix}
    1 & 0& 0 & 0 \\
   0 & 0&1 &0 \\
    0& 1 &0 &0 \\
 0 & 0 & 0 & 1
\end{pmatrix}
 \otimes I  \right].
 \end{align}
 Note that $I\in\mathbb{R}^{2\times2}$ and, hence, $P\in\mathbb{R}^{16\times16}$.

\section{Noise Analysis}\label{app:noise}
\renewcommand{\theequation}{D\arabic{equation}}
\setcounter{equation}{0}

We follow the analysis of~\cite{li2022single}. The equations that we used to simulate the various noise processes are detailed below. 

\subsection*{I. Doppler shifts and fluctuations in atomic separation}

Finite temperature effects give rise to fluctuations in atomic  velocity and position. The former produce shifts in the effective single-photon detuning  while the latter introduce a variation in the Rydberg interaction strength. To see how this comes about, let us write the Hamiltonian.
We denote  atomic positions by $\vec{R}_\ell(t) = \vec{R}_\ell + \delta\vec{R}_\ell + \vec{v}_\ell t$, where the index relates to the $\ell^{\mathrm{th}}$ atom, $\delta\vec{R}_\ell$ are position deviations and $\vec{v}_\ell$ are the atomic velocities.  The Hamiltonian is:
\begin{widetext}
    \begin{equation}
\Scale[0.95]{H = \sum_\ell \left\{\frac{\Omega_1(t)}{2}e^{i\vec{k}_1\cdot\vec{R}_\ell(t)}\KET{1_\ell}\BRA{p_\ell} + 
\frac{\Omega_2(t)}{2}e^{i\vec{k}_2\cdot\vec{R}_\ell(t)}\KET{p_\ell}\BRA{r_\ell} +\mbox{h.c.}- 
\Delta\KET{p_\ell}\BRA{p_\ell}\right\} + V_R(|\vec{R}_1-\vec{R}_2|)\KET{r_1r_2}\BRA{r_1r_2}}
\end{equation}
\end{widetext}

where $\vec{k}_j$ is  the wavevector of the control pulses $\Omega_j$.  We assume in our analysis that the control pulses and the tweezer beams  are co-propagating along the $z$ direction (i.e., the wavevectors are parallel). The atoms are placed along an axis perpendicular to $z$
(say along the $x$ direction). Using the parameters from~\cite{levine2022dispersive}, we consider a temperature of 10 $  \,\mu$K which produces a normal  distribution of detunings with standard deviation $2\pi\times43$  kHz. Assuming tightly focused tweezer beams,  we a distribution of positions with standard deviation of $200$ nm.

\subsection*{II. Spatial dependence of the pulse amplitude}

The control pulses are  Gaussian beams. 
For simplicity, we consider the effect of amplitude fluctuations due to atom deviation from the trap center in the transverse direction (x,y) only. 
Hence, the amplitude of control pulse $j = 1,2$ depends on the transverse (x,y) coordinates of atom $\ell$ via the relation:
\begin{equation}
\Omega_j(t,\vec{R}_\ell) =  \Omega_j(t,0)e^{(x_\ell^2+y_\ell^2)/w^2}
\end{equation}
where $w$ is the beam waist, chosen here to be $w = 1  \,\mu$m. We generate a random distribution of locations corresponding to a temperature of 10 $  \,\mu$K, using trap parameters from~\cite{li2022single}.

\subsection*{III. Noise in the amplitude and phase of the control fields}

Classical noise in the amplitude of the control fields is accounted for by generating an ensemble of amplitude deviations $\Omega_j\rightarrow\Omega_j+\delta_j$ and averaging over the resulting fidelity. Previous work has shown that this noise channel has little effect on STIRAP-based protocols and we confirm this result. Laser phase noise has an average effect of introducing a dephasing channel, which is conveniently modeled by an appropriate jump term in the Lindbladian. Specifically, following~\cite{li2022single}, we include a term of the form $\sqrt{\gamma_1/2}(\KET{p}\BRA{p}-\KET{1}\BRA{1})$ and 
$\sqrt{\gamma_2/2}(\KET{r}\BRA{r}-\KET{p}\BRA{p})$ to account for the noise in each control field. 

\bibliographystyle{sn-aps}
\bibliography{STIRAPbibli}

\end{document}